\def\e{e}
\def\g{\sigma}
\def\q{{\bf q}}
\def\p{{\bf p}}
\def\h{{\bf h}}
\def\v{{\bf v}}
\def\bzeta{\mbox{\boldmath$\zeta$}}
\def\bxi{\mbox{\boldmath$\xi$}}
\def\tbxi{\mbox{\boldmath$\tilde\xi$}}
\def\grad{\mbox{\boldmath$\nabla$}}

\def\P{{\cal P}}
\def\tr{{\rm tr}}
\def\half{{\textstyle{1\over2}}}
\def\T{{\cal T}}
\def\W{_{\rm W}^{}}

\def\F{{\bf F}}
\def\bFito{\F^{\mbox{\tiny It\^o}}}
\def\Fito{F^{\mbox{\tiny It\^o}}}
\def\bFstrat{\F^{\mbox{\tiny Strat}}}
\def\Fstrat{F^{\mbox{\tiny Strat}}}
\def\Peq{P_{\rm eq}}

\documentstyle[preprint,aps,fixes,epsfig]{revtex}
\tightenlines
\begin {document}



\preprint {UVA/Arnold--99--44}
\title
{Langevin equations with multiplicative noise: resolution of
time discretization ambiguities for equilibrium systems}

\author {Peter Arnold}

\address
    {%
    Department of Physics,
    University of Virginia,
    Charlottesville, VA 22901
    }%
\date {November 1999}

\maketitle
\vskip -20pt

\begin {abstract}%
{%
A Langevin equation with multiplicative noise is an equation schematically
of the form $d\q/dt = -\F(\q) + e(\q) \, \bxi$, where $e(\q)\,\bxi$ is Gaussian
white noise whose amplitude $e(\q)$ depends on $\q$ itself.
Such equations are ambiguous, and depend on the details of one's convention
for discretizing time when solving them.  I show that these ambiguities are
uniquely resolved if the system has a known equilibrium distribution
$\exp[-V(\q)/T]$ and if, at some more fundamental level, the physics of
the system is reversible.
I also discuss a simple example where this
happens, which is the small frequency limit of Newton's equation
$\ddot\q + e^2(\q) \, \dot\q = - \grad V(\q) + e^{-1}(\q) \, \bxi$ with
noise and
a $\q$-dependent damping term.
The resolution does {\it not}\/ correspond to simply interpreting naive
continuum equations in a standard convention, such as
Stratanovich or It\^o.
}%
\end {abstract}

\thispagestyle{empty}


\section {Introduction}

A Langevin equation of the form
\begin {mathletters}
\label{eq:theory0}
\begin {equation}
   \sigma \dot\q = - \grad V(\q) + \bzeta ,
\label{eq:L0}
\end {equation}
\begin {equation}
   \langle \zeta_i(t)\,\zeta_j(t') \rangle
   = 2\sigma T \, \delta_{ij} \, \delta(t-t') ,
\end {equation}
\end {mathletters}%
can appear as the effective description of highly overdamped
motion of some coordinates
$\q$ in a potential $V$ in contact with a thermal bath.
The damping, and the random force term, arise from interactions with the
thermal bath.
By overdamped, I mean that at a slightly more fundamental level, the
equation of motion might be, for instance,
\begin {equation}
   \ddot\q + \sigma \dot\q = - \grad V(\q) + \bzeta ,
\label{eq:Newt1}
\end {equation}
with $\bzeta$ as before.  Here, $\sigma$ is a damping coefficient,
and $\bzeta$ is a Gaussian thermal noise term.
In the limit that $\sigma$ is large compared to the inverse time scales
of interest, one can ignore the $\ddot\q$ term and so obtain
(\ref{eq:L0}).

In this article, I want to consider the case where (a) the damping
coefficient $\sigma$ depends on $\q$ itself,
\begin {mathletters}
\label {eq:L1}
\begin {equation}
   \sigma_{ij}(\q) \, \dot q_i = - \nabla_i V(\q) + \zeta_i ,
\end {equation}
\begin {equation}
   \langle \zeta_i(t)\,\zeta_j(t') \rangle
   = 2\sigma_{ij}(\q)\, T \, \delta(t-t') ,
\label{eq:L1b}
\end {equation}
\end {mathletters}%
and (b) it is already know through other means that
$V(\q)$ is an effective potential that gives the equilibrium
distribution of $\q$
as
\begin {equation}
   \Peq(\q) = \exp[-V(\q)/T]
\label{eq:Peq}
\end {equation}
in whatever approximation may be relevant to the problem of interest.
Equation (\ref{eq:L1}) is a special case of what are known as Langevin
equations with multiplicative noise.
In general, such equations are notorious for being ambiguous---they
are sensitive to exactly how one discretizes time.
The purpose of this paper is to show that, with a few very general
assumptions, the property (\ref{eq:Peq}) is sufficient to uniquely
resolve this ambiguity.  Moreover, this resolution does {\it not}\/
correspond to any standard interpretation (It\^o or Stratanovich)
of the Langevin equation (\ref{eq:L1}).

Though my arguments will be somewhat general, it helps to have in mind
a concrete example of an {\it un}-ambiguous description for which
(\ref{eq:L1}) will be a limiting case.  A simple example is
the analog of (\ref{eq:Newt1}) with $\q$-dependent damping:
\begin {equation}
   \ddot q_i + \sigma_{ij}(\q) \, \dot q_j = - \nabla_i V(\q) + \zeta_i ,
\label{eq:LangInertia}
\end {equation}
with the noise again (\ref{eq:L1b}).
As I discuss in Appendix \ref{app:inertia},
this equation is unambiguous and has
equilibrium distribution (\ref{eq:Peq}).

I can't resist mentioning the application of interest to me personally,
which motivated this work \cite{arnold}:
electroweak baryon number violation
in the early universe.
Its study requires understanding the effective dynamics
of fluctuations in weakly-coupled high-temperature non-Abelian gauge theories,
where there is an effective theory of the form (\ref{eq:L1}) at the relevant
distance and time scales, but where it is much more straightforward to
analyze static issues, such as (\ref{eq:Peq}), than the subtleties of
dynamical ones.


\section {A first pass}
\label{sec:first}

Let's rewrite (\ref{eq:L1}) in the equivalent form
\begin {mathletters}
\label {eq:G1}
\begin {equation}
   \dot \q = - \F(\q) + \e(\q) \, \bxi_i ,
\end {equation}
\begin {equation}
   \langle \xi_i(t)\,\xi_j(t') \rangle = 2T \, \delta_{ij} \, \delta(t-t') .
\end {equation}
\end {mathletters}%
where
the $\q$ dependence has been scaled out of the noise by defining
$\bxi \equiv \e(\q) \, \bzeta$, and the matrix $\e$ and vector $\F$ are
\begin {equation}
   \e(\q) \equiv [\sigma(\q)]^{-1/2} ,
\end {equation}
\begin {equation}
   \F(\q) \equiv \g^{-1}(\q) \, \grad V(\q) .
\label{eq:Fdef}
\end {equation}
Two standard conventions for discretizing time, and so removing the
ambiguities inherent in equations like (\ref{eq:L1}), are the
It\^o convention,%
\footnote{
   For a general review of background material for this paper, in notation
   close to what I use here, see, for example, chapter 4 of ref.\ 
   \cite{Zinn-Justin}.  The most substantial differences in notation are
   that my $\F$ and $T$ are that reference's ${1\over2} {\bf f}$
   and ${1\over2}\Omega$.
}
\begin {equation}
   \q_t - \q_{t-\Delta t}
   = - \Delta t \, \bFito(\q_{t-\Delta t}) + e(\q_{t-\Delta t}) \,
           \tbxi_t ,
\label {eq:LangIto}
\end {equation}
and the Stratanovich convention,
\begin {equation}
   \q_t - \q_{t-\Delta t}
   = - \Delta t \, \bFstrat(\bar\q) + e(\bar\q) \, \tbxi_t ,
\label {eq:LangStrat}
\end {equation}
\begin {equation}
   \bar\q \equiv {\q_t+\q_{t-\Delta t} \over 2} \,,
\end {equation}
where in both cases the discretized noise correlation is
\begin {equation}
   \langle \tilde\xi_{it} \tilde\xi_{jt'} \rangle
   = 2 T \, \Delta t\, \delta_{ij} \delta_{t t'} .
\end {equation}
The specification of $\bFstrat(\bar\q)$ as opposed to
$\bFstrat(\q_{t-\Delta t})$ in (\ref{eq:LangStrat}) is actually irrelevant:
it is only the $\q$ used to evaluate $e$ that causes the difference
between these two conventions in the $\Delta t \to 0$ limit.

If one simply tries setting $\bFito = \F$ or $\bFstrat = \F$, the
It\^o and Stratanovich equations will
give rise to different physics, and in particular
different equilibrium distributions.  The two conventions are
identical in the $\Delta t \to 0$ limit only if one sets
\begin {equation}
   \Fito_i = \Fstrat_i - T \e_{ia,j} \e_{ja} .
\end {equation}
(Here and throughout, I adopt the notation that indices after a comma
represent derivatives: $F_{i,j} \equiv \partial F_i/\partial q_j$
and $F_{i,jk} \equiv \partial^2 F_i/\partial q_j \partial q_k$.)

The value of $\F$ that should be used for a particular physical problem
therefore depends on what discretization convention one picks to use.
Since the value of $\F$ is ambiguous, one approach might be to
simply pick a discretization convention (it doesn't matter which one),
and then choose $\F$ however necessary to reproduce the desired
equilibrium distribution (\ref{eq:Peq}).  To be concrete, let's
choose Stratanovich convention.  The Stratanovich Langevin equation
(\ref{eq:LangStrat}) is well known to be equivalent to the
following Fokker-Planck equation for the time evolution of the
probability distribution $P(\q,t)$:
\begin {equation}
   \dot P = \nabla_i \left[ T \e_{ia} \nabla_j (\e_{ja} P)
                                     + \Fstrat_i P \right] .
\label{eq:FP}
\end {equation}
My requirement is that this equation have the equilibrium distribution
(\ref{eq:Peq}) as a time-independent solution.  A simple way to achieve this
is for
\begin {equation}
   T \e_{ia} \nabla_j (\e_{ja} \Peq) + \Fstrat_i \Peq = 0,
\end {equation}
which gives
\begin {equation}
   \Fstrat_i = (\g^{-1})_{ij} \nabla_j V - T \e_{ia} \e_{ja,j} .
\label{eq:Fstrat}
\end {equation}
Compare to our naive starting point (\ref{eq:Fdef}).
Equivalently,
\begin {equation}
   \Fito_i = (\g^{-1})_{ij} \nabla_j V - T (\g^{-1})_{ij,j} \,.
\label {eq:Fito}
\end {equation}
So one might suspect that the Stratanovich equation (\ref{eq:LangStrat})
together with (\ref{eq:Fstrat}), or equivalently the It\^o equation
(\ref{eq:LangIto}) together with (\ref{eq:Fito}), gives the correct
description of the system.  As we shall see, this is indeed the case.

Based on the presentation so far, the reader might be suspicious of two
things.  First, once we give ourselves license to go mucking around
with the equation, changing what we thought was $\F(\q)$ to suit our needs,
how do we know we aren't supposed to change other things as well?
In particular, what tells us that we shouldn't change
$e(\q)$ in some way,
then make some compensating change in $\F$ to force the equilibrium
distribution to work out?
The basic answer is that $\F$ is sensitive to the details of short-time
physics and regularization, whereas $e$ is not, as evidenced by the fact
$\F$ must be changed when one adopts different conventions like
It\^o or Stratanovich, but $e$ need not.
This boils down to a discussion of the renormalizeability of the theory,
and its consequences for how, in principle, the theory
should be matched to a more
fundamental description of the physics.  Such issues are more familiar
in the context of theories defined by path integrals, and so
much of the rest of this paper will be to translate the discussion into
that language.

Secondly, the Stratanovich $\F$ of (\ref{eq:Fstrat}) is not the unique
solution to (\ref{eq:FP}) for $P = \Peq$.  The general solution is
\begin {equation}
   \Fstrat_i = (\g^{-1})_{ij} \nabla_j V - T \e_{ia} \e_{ja,j}
                      + h_i e^{+V/T} ,
\label {eq:Fh}
\end {equation}
where $\h = \h(\q)$ is any function with $\grad\cdot\h = 0$.

I shall throughout focus on systems where the underlying physics,
whatever it may be, is reversible.  More specifically, I shall assume that
the effective theory (\ref{eq:theory0}) must be defined in such a way that
equilibrium time-dependent correlations, such as
$\langle \q(t) \, \q(0) \rangle$, are invariant under time reversal.
[As discussed in Appendix \ref{app:inertia},
the behavior of the Langevin equation with
inertia (\ref{eq:LangInertia}) has this property, even though the equation
by itself is not time-reversal invariant.]
As I'll show, this assumption will rule out the extra term involving $\h$ in
(\ref{eq:Fh}).
I shall then discuss in more detail, in the language of path integrals,
how the issue of whether
$e$ or $\F$ should be
modified from their ``naive'' values is an issue of renormalizeability.
Specifically, I will define ``naive'' by assuming that
there is some more fundamental description
of the effective theory, such as the Langevin equation (\ref{eq:LangInertia})
with inertia, that is unambiguous and is described in terms of the same
degrees of freedom $\q$.  The ``naive'' values of $e$ and $\F$ will simply
be those defined by naively taking the low frequency limit of the
more fundamental equation.
I'll show that terms associated with $e$ require no
ultraviolet renormalization, while terms associated with $\F$ do.
Based on general procedures for matching effective theories to more
fundamental underlying theories, one may then argue that $\F$, and not
$e$, should be modified to make the physics work out right.
This is something that, I believe, may be well known to the few people
to whom it is well known.  However, since there seems to be general
confusion on this matter, it seems worthwhile to continue rather than
simply ending here.


\section {The path integral version}

The path integral corresponding to the Stratanovich Langevin equation
(\ref{eq:LangStrat}) is, somewhat imprecisely, \cite{path}%
\footnote{
   There is a small change of notation from ref. \cite{path}: the
   $(\det e)^{-1}$ pre-exponential factor of that reference has been 
   moved up into the exponent and absorbed into $L$.
}%
$^{,}$%
\footnote{
  The formalism here is somewhat similar to that used to describe
  stochastic processes on curved manifolds, if one were to interpret
  $\sigma_{ij}$ as a metric tensor $g_{ij}$.  I should emphasize that
  this is {\it not}\/ the problem I am studying.
  On a curved manifold, the desired equilibrium distribution would be
  $[\det \g]^{1/2} \exp(-V/T)$ instead of $\exp(-V/T)$.
  Mathematically, one can convert between these two problems by
  setting $V_{\rm manifold} = V + \half T \ln \det \g$.
}
\begin {eqnarray}
   \P(\q'',\q',t''-t')
   &=& \int_{\q(t')=\q'}^{\q(t'')=\q''}
       \left[d\q(t)\right] \>
      \exp\left[ - \int_{t'}^{t''} dt\>
          L(\dot\q,\q)
      \right] ,
\label{eq:path}
\end {eqnarray}
\begin {equation}
   L(\dot\q, \q) = {1\over 4T} \, (\dot q + F)_i \g_{ij} (\dot q + F)_j
      - {1\over2} \, F_{i,i}
      + {1\over2} \, e^{-1}_{ia} e_{ka,k}(\dot q + F)_i
      + {T\over4} \, e_{ia,j} e_{ja,i}
      + \delta(0) \, \tr \ln e .
\label{eq:Lresult}
\end {equation}
Here and henceforth, I abbreviate $\bFstrat$ as simply $\F$.
The imprecision is just due to the fact that this path integral depends on
the details of how time is discretized.  I've implicitly
assumed a time-symmetric
discretization above.  Specifically, (\ref{eq:path}) really means
\begin {eqnarray}
   \P(\q'',\q',t''-t')
   &=& \lim_{\Delta t \to 0} N
      \int_{\q(t')=\q'}^{\q(t'')=\q''}
      \left[\prod_t d\q_t \right] \>
\nonumber\\ && \qquad\qquad \times
      \exp\left[ - \Delta t \sum_t
          L\left( {\q_t - \q_{t-\Delta t} \over \Delta t} \, , \,
                  {\q_t + \q_{t-\Delta t} \over 2} \right)
      \right] ,
\label {eq:discretepath}
\end {eqnarray}
where $N$ is an overall normalization I shan't bother being explicit about,
and where $\delta(0)$ in (\ref{eq:Lresult}) means $(\Delta t)^{-1}$.
If I had used some other discretization convention in the path integral,
the Lagrangian $L$ would be different from (\ref{eq:Lresult}).

For my argument, it will be sufficient (and convenient) to focus on
fluctuations about equilibrium.  It is then enough to consider
the $t' \to -\infty$ and $t'' \to +\infty$ limit of the path integral, writing
\begin {equation}
   Z
   \equiv \int
       \left[d\q(t)\right] \>
      \exp\left[ - \int_{-\infty}^{+\infty} dt\>
          L(\dot\q,\q)
      \right] .
\label{eq:Z}
\end {equation}
Because the system is dissipative, the boundary conditions on $\q$ at
$t=\pm\infty$ decouple.
Equilibrium correlation functions
like $\langle \q(t) \q(0) \rangle$ can then be
evaluated in the usual way as
\begin {equation}
   \langle \cdots \rangle
   = Z^{-1} \int
       \left[d\q(t)\right] \>
      \exp\left[ - \int_{-\infty}^{+\infty} dt\>
          L(\dot\q,\q)
      \right] \cdots .
\end {equation}

This form allows us to rewrite the Lagrangian in a form where the
assumed time-reversal symmetry of correlations is manifest.
Simply as an illustrative example, suppose for the moment that $\F$
{\it did}\/ have the
form (\ref{eq:Fstrat}) that I'm trying to demonstrate.
The Lagrangian (\ref{eq:Lresult}) is not manifestly invariant under
$t \to -t$.
However, for the choice (\ref{eq:Fstrat}) of $\F$, this Lagrangian can
be rewritten as
\begin {eqnarray}
   L(\dot\q, \q) &=&
      {1\over 4T} \, \dot q_i \g_{ij} \dot q_j
      + {1\over 4T} \, (\nabla_i V) (\g^{-1})_{ij} (\nabla_j V)
      - {1\over2} \nabla_i [(\g^{-1})_{ij} \nabla_j V]
      + {T\over4} (\g^{-1})_{ij,ij}
\nonumber\\ && \hspace{8em}
      + \delta(0) \, \tr \ln e
      + {1\over 2T} \, \partial_t V .
\label {eq:LV}
\end {eqnarray}
Only the last term is not manifestly time-reversal invariant.
However, with the path integral in the current form of (\ref{eq:Z}), we are
allowed to throw away terms in $L$ that are total time derivatives.
That's because they can be integrated
and re-expressed in terms of the boundary conditions, which we know are
irrelevant.  [As reviewed in Appendix \ref{app:parts},
the fact that one may naively
integrate total time derivatives is dependent upon the use of the symmetric
time discretization (\ref{eq:discretepath}).]
The upshot is, that for the purpose of computing equilibrium
(but time-dependent) correlators, we can replace $L$ by
\begin {eqnarray}
   L_2(\dot\q, \q) &=&
      {1\over 4T} \, \dot q_i \g_{ij} \dot q_j
      + {1\over 4T} \, (\nabla_i V) (\g^{-1})_{ij} (\nabla_j V)
      - {1\over2} \nabla_i [(\g^{-1})_{ij} \nabla_j V]
      + {T\over4} (\g^{-1})_{ij,ij}
\nonumber\\ && \hspace{8em}
      + \delta(0) \, \tr \, \ln e \,.
\label {eq:L2}
\end {eqnarray}
It's important to emphasize that $L_2$ cannot be used in the finite-time
path integral (\ref{eq:path}): it would not produce the same
${\cal P}(\q'',\q',t''-t')$.

In passing, we can now see the problem if
$\F$ contained the
additional $h$ term of (\ref{eq:Fh}).  The Lagrangian $L$ given by
(\ref{eq:Lresult}) would then produce additional terms in (\ref{eq:LV}),
one of which,
\begin {equation}
   {1\over 2T} \, \dot q_i \g_{ij} h_j e^{V/T} ,
\label{eq:hprob}
\end {equation}
is odd under time reversal.
So the appearance of $h$ would contradict my assumption about reversibility,
unless (\ref{eq:hprob}) can be waived away as a total time derivative.
This would require
\begin {equation}
   \delta F_i \equiv h_i e^{V/T} = (\g^{-1})_{ij} \nabla_j \phi
\end {equation}
for some function $\phi(\q)$.  But, returning to (\ref{eq:Fstrat}),
this just corresponds to a shift of $V \to V + \phi$, and the equilibrium
distribution produced by the Fokker-Planck equation (\ref{eq:FP}) would
then be $\exp[-(V+\phi)/T]$ instead of the required $\exp(-V/T)$.

Now let's divorce ourselves from the particular form (\ref{eq:Fstrat})
that I've claimed for $\F$.  Instead, just write down the
most general form for a manifestly time-reversal invariant $L_2$:
\begin {equation}
   L_2(\dot\q, \q) =
      {1\over 4T} \, \dot q_i \, \g_{ij}(\q) \, \dot q_j
      + U(\q) .
\label {eq:Lgeneric}
\end {equation}
Before discussing how to determine the correct choices of
$\g_{ij}$ and $U$ for this equation, I want to discuss its
renormalization.%
\footnote{
   If the discussion
   in this paper is applied to a field theory, then I am assuming that
   the theory has already been regulated spatially, by the introduction
   of a small distance cut-off (either explicitly, by a spatial lattice
   cut-off, or implicitly by, for example, dimensional regularization),
   and that I am now separately considering the issue of regularizing
   and renormalizing small-time divergences.
   The discussion of divergences and power counting will therefore be
   somewhat different from what is usual in field theory, where one
   typically regulates space and time simultaneously.
}
Imagine that the Lagrangian could be expanded in perturbation theory
about some $\q_0$ with $\g(\q_0)$ finite and non-zero.
For simplicity, say $\q_0 = 0$.
The arguments that follow will be about the entire
perturbative expansion, to all orders in perturbation theory.
To this end, write
\begin {equation}
   {1\over 4T} \, \dot q_i \, \g_{ij}(\q) \, \dot q_j =
   {1\over 4T} \left[ \dot q_i \, \g_{ij}(0) \, \dot q_j +
           \dot q_i \, \delta \g_{ij}(\q) \, \dot q_j \right] ,
\end {equation}
and formally consider $\delta \g$ as a perturbation.

\begin {figure}[th]
\vbox{
   \begin {center}
      \epsfig{file=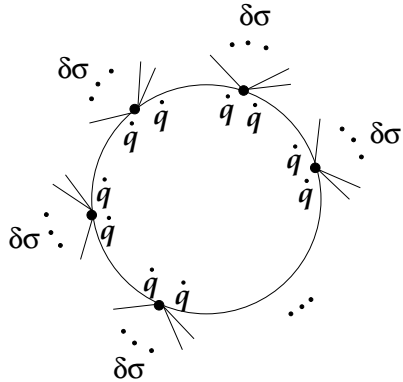,scale=.6}
   \end {center}
   \caption{
       The UV divergent 1-loop graphs.
       \label{fig:1loop}
   }
}
\end {figure}

I will now show that the $U(\q)$ term in (\ref{eq:Lgeneric})
requires counter-terms,
but the $\g(\q)$ term does not.
The Lagrangian (\ref{eq:L2}) is super-renormalizeable: counter-terms
are needed only for one loop and two loop diagrams.
This can be seen by simple power counting of diagrams.
The most ultraviolet (UV) divergent diagrams will be those whose
vertices involve as many powers as possible of internal frequency.
That is, vertices generated by the $\dot q_i \, \delta \g_{ij}(\q) \dot q_j$
terms in (\ref{eq:L2}) will give more UV divergent behavior than vertices
generated by the $U(\q)$ terms.  And the most divergent behavior will occur
when the $\dot\q$'s are associated with internal rather than external lines.
The most divergent one-loop diagrams are therefore those depicted by
fig.\ \ref{fig:1loop}.  The UV behavior of such graphs is
\begin {equation}
   \int d\omega\> {(\omega^2)^V \over (\omega^2)^I} = \int d\omega ,
\label {eq:1power}
\end {equation}
where $(\omega^2)^V$ are the powers of $\omega$ from the $V$ vertices
$\dot\q \, \delta \g \, \dot\q$, and $(\omega^2)^{-I}$ are the
$\omega \to\infty$ behavior of the $I=V$ internal propagators.
So this class of diagrams is linearly divergent in the UV.
The linearly divergent piece is independent of external frequencies,
and so requires a counterterm in $U(\q)$ but not one in the
kinetic term $\dot\q \g \dot\q$.
That is, so far we have seen that the effective $U(\q)$ is dependent
on the details of short-time physics, but we have not seen any
evidence that $\g(\q)$ is.

Are there divergences associated with external frequency dependence?
Expanding the diagrams of fig.\ \ref{fig:1loop} in powers of external
momenta, each such power will come at the expense of one of the
factors of $\omega$ in (\ref{eq:1power}).  So no counter-terms
involving two external frequencies are necessary---that is, no
counterterm for $\dot\q \g \dot\q$ is necessary.
There could be a log divergence associated with one external frequency
factor, except that the resulting $\omega$ integral $\int d\omega/\omega$
vanishes by the assumption that the equilibrium physics is time-reversal
invariant (and the fact that I am using a symmetric discretization of the
path integral, which respects that invariance).

Similarly, one could imagine a different method of hooking up the
diagrams of fig.\ \ref{fig:1loop} in which one or more of the
$\dot\q$ factors were associated with external rather than internal
lines.  If this happens for two or more factors, the loop integral is
no longer UV divergent and so in not sensitive to short-time physics.
If it happens for only one factor, then the loop integral again vanishes
by the assumption of time-reversal invariance.
Therefore, the linear UV divergence (\ref{eq:1power})
previously discussed is the
{\it only}\/ UV divergence at one loop.

\begin {figure}[th]
\vbox{
   \begin {center}
      \epsfig{file=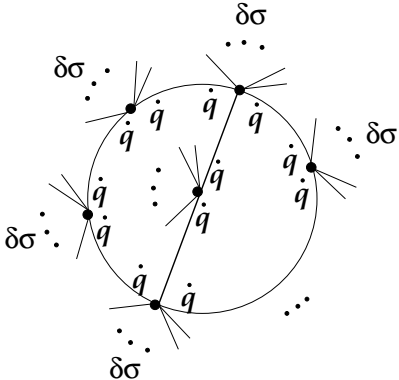,scale=.6}
   \end {center}
   \caption{
       The UV divergent 2-loop graphs.
       In the vertices with three internal lines, the $\dot\q$'s can be
       associated with any two of them.
       \label{fig:2loop}
   }
}
\end {figure}

Now consider two loop diagrams, again maximizing the divergence by
using $\dot\q \, \delta \g \, \dot\q$ vertices rather than $U(\q)$ ones,
and associating all $\dot\q$'s with internal lines.
The result is shown in fig.\ \ref{fig:2loop}.  Simple power counting
gives the maximum possible new divergence at two loops as order
\begin {equation}
   \int d^2\omega\> {(\omega^2)^V \over (\omega^2)^I}
   = \int {d^2\omega \over \omega^2} ,
\label {eq:2power}
\end {equation}
where $I = V+1$ and $\omega$ just counts powers of combinations of
the two loop frequencies $\omega_1$ and $\omega_2$.
A logarithmic divergence is possible, but this divergence has no
dependence on external frequencies.  As before, if we focus on the
dependence of these diagrams on external frequency, then there is
no sensitivity to short-time physics.  If we go to yet higher orders
in diagrams, there are no new divergences at all: naive power counting
gives $\int \omega^{-2(L-1)} \> d^L\omega$ for an $L$-loop diagram.

The upshot of all this is that only $U(\q)$ is sensitive to the details
of short-time physics; $\dot\q \g \dot\q$ is not.

In the language of path integrals, the ``naive'' low frequency limit of
a theory is what you get if you simply replace the Lagrangian by its
low frequency limit.  This replacement is naive when there are UV
divergences in the resulting effective theory, which make the physics
of that theory sensitive to how it is cut off in the UV---that is,
sensitive to the details of the more fundamental theory.
Fortunately, these details can be absorbed into a redefinition of
the effective interactions which depend on them.  In this case, that
means $U(\q)$ is modified from its naive form, but $\dot\q \g \dot\q$
is not.

Since this point is important, let me state it another way.
The basic theory and technology of matching parameters of low-frequency
effective theories to those of more fundamental theories, to any and all
orders in perturbation theory, has been
used in a number of problems in field theory,
including Bose condensation \cite{Bose},
ultrarelativistic plasmas \cite{Braaten&Nieto},
heavy quark physics \cite{heavy quarks},
and non-relativistic plasma physics \cite{nonrel plasma}.
(For a general discussion, see also ref.\ \cite{effective}.)
The basic idea is to fix a regularization for the effective theory, to
then treat all the parameters of the effective theory
as adjustable, to calculate low-frequency observables in both the effective
theory and the more fundamental theory, and then to fix the parameters
of the effective theory to obtain agreement.
If we simply fixed $\g(\q)$ and $U(\q)$ to be their naive values, everything
would match order by order in perturbation theory if it weren't for the
UV divergences in the effective theory---if diagrams only involved
low frequencies in propagators and vertices, then each internal line
and vertex in the underlying theory would match up with each one in the
naive effective theory.  The only thing we have to correct for are UV
divergent graphs or sub-graphs, where the correspondence would fail.
But the previous discussion shows that such divergences have the form
of counter-terms for $U(\q)$, and so the matching of the two theories can
be fixed by appropriate adjustment of those counter-terms.

If we had a specific underlying theory in mind, we cold carry out this
matching procedure by computing to two loops in perturbation theory.
The crucial
point here is that such a calculation is unnecessary, and detailed knowledge
of the underlying theory is unnecessary, if we know that the
equilibrium distribution is (\ref{eq:Peq}).  We can then immediately solve
the problem by simply requiring that we choose $U(\q)$ so that the
equilibrium distribution comes out right.

Returning to the Lagrangian (\ref{eq:Lresult}) corresponding to the original
Langevin equation, the only way we can change $U(\q)$ without changing
$\dot\q \g \dot\q$ is by changing $\F$, which justifies the argument
for choosing $\bFstrat$ given in section \ref{sec:first}.
Alternatively, one can uniquely determine $U(\q)$ directly by
requiring that the Euclidean Schr\"odinger equation, corresponding to
the path integral with the generic Lagrangian $L_2$ (\ref{eq:Lgeneric}),
generate $\exp(-V/T)$ as its equilibrium distribution.
This is carried out in Appendix \ref{app:H}, which also explains
some interesting distinctions between the interpretation of the
Euclidean Schr\"odinger equation associated with $L_2$
[(\ref{eq:L2}) or (\ref{eq:Lgeneric})] and
that associated with $L$ (\ref{eq:Lresult}).
The final result agrees with the initial analysis of section (\ref{sec:first}).


\section* {ACKNOWLEDGMENTS}

I thank Larry Yaffe, Dam Son, and Tim Newman for useful conversations.  This
work was supported by the U.S. Department of Energy under Grant No.\
DEFG02-97ER41027.  I thank the Department of Energy's Institute for Nuclear
Theory at the University of Washington for its hospitality during completion
of this work.


\appendix

\section{The Langevin equation with inertia}
\label{app:inertia}

The Langevin equation (\ref{eq:LangInertia}) with inertia can be
written in the form of a standard first-order Langevin equation
by rewriting it as a system of first order equations, introducing
$\v \equiv \dot\q$:%
\footnote{
   Interestingly, this is a very special case of a general set of
   Langevin equations considered in ref.\ \cite{Ramshaw}.
}
\begin {mathletters}
\label{eq:inertia2}
\begin {eqnarray}
   \dot\v + \sigma(\q) \,\v &=& - \grad V(\q) + e^{-1}(\q) \, \bxi ,
\label{eq:inertia2a}
\\
   \dot\q &=& \v .
\end {eqnarray}
\end {mathletters}%
This equation is free of the usual discretization ambiguities.
To see this, consider the difference between Stratanovich and It\^o
discretization conventions, which differ in how the $\q$ in the
$e^{-1}(\q) \, \bxi$ term is evaluated.  That difference in $\q$ is
${1\over2} \dot\q \, \Delta t$, which means a difference in
the $e^{-1}(\q) \, \bxi$ term of
\begin {equation}
   {\Delta t\over2} \, (e^{-1})_{ij,k} \dot q_k \xi_j .
\label{eq:ediff}
\end {equation}
In a Langevin equation such as our effective theory
(\ref{eq:G1}), this would not vanish in the $\Delta t \to 0$ limit
because $\bxi$ is order $(\Delta t)^{-1/2}$, and then so is $\dot\q$
by the equation of motion (\ref{eq:G1}).%
\footnote{
  See, for example, secs.\ 4.7--8 of ref.\ \cite{Zinn-Justin} for a review.
}
In the case at hand, matters are different, because $\dot q_k = v_k$,
and, though (\ref{eq:inertia2a}) implies that the time derivative
$\dot\v$ will be order $(\Delta t)^{-1/2}$, $\v$ itself is perfectly
finite as $\Delta t \to 0$.  So, in the case at hand, the difference
(\ref{eq:ediff}) vanishes in the $\Delta t \to 0$ limit.

It is also interesting to sketch how this works in the
path integral formulation.
Converting equation (\ref{eq:LangInertia}) directly into a path integral
using standard methods%
\footnote{
   See, for example, sec.\ 4.6 of ref.\ \cite{Zinn-Justin} for a review,
   as well as ref.\ \cite{path}.
}
gives a path integral with Lagrangian
\begin {equation}
   L = {1\over 4T} \, (\ddot\q + \g \dot\q + \grad V) \g^{-1} 
                      (\ddot\q + \g \dot\q + \grad V)
       + L_\eta
       + \delta(0) \, \tr \ln e ,
\label {eq:Linertia}
\end {equation}
\begin {equation}
   L_\eta = \bar\eta_i \left\{
       \delta_{ij} \partial_t^2
       + \g_{ij} \partial_t
       + (e^{-1})_{ai} \left[
             e_{ka,j} \ddot q_k
             + (e^{-1})_{ak,j} \dot q_k
             + \nabla_j (e_{ka} \nabla_k V)
         \right]
       \right\} \eta_j ,
\end {equation}
where $\bar\eta$ and $\eta$ are anti-commuting Grassman fields (ghosts).
A diagrammatic analysis similar to that discussed in the main text produces
no UV divergences because the $\ddot\q \g^{-1} \ddot\q$ term means that
$\q$ propagators go like $\omega^{-4}$ in the UV instead of only
$\omega^{-2}$.  So diagrams converge faster in the UV than they did in
the previous analysis.

Now let's check the equilibrium distribution for $\q$.  Since
(\ref{eq:inertia2}) formally has the same form as a generic first-order
Langevin equation (\ref{eq:G1}) [the number of degrees of freedom have simply
doubled from $\q$ to $(\q,\v)$], we can use the result (\ref{eq:FP})
for the Fokker-Planck equation for the probability distribution
$P(\q,\v,t)$, which translates in the present case to
\begin {eqnarray}
   \dot P &=& \nabla_{v_i}
              \left\{ T (e^{-1})_{ia} \nabla_{v_j} [(e^{-1})_{ja} P]
                + (\sigma\v + \grad_\q V)_i P \right\}
          + \nabla_{q_i} \left\{ - v_i P \right\}
\nonumber\\
   &=& \sigma_{ij} \left[ T \nabla_{v_i} \nabla_{v_j} P
             + \nabla_{v_i} \cdot (v_j P) \right]
       + \grad_\q V \cdot \grad_\v P
       - \v\cdot\grad_{\q} P .
\end {eqnarray} 
The equilibrium solution is
\begin {equation}
   P_{\rm eq}(\dot\q,\q) =
   \exp\left\{ - {1\over T}\left[\half v^2 + V(\q) \right] \right\} .
\end {equation}
If we are only interested in the distribution in $\q$ and not $\v=\dot\q$,
we may integrate over all possible values of $\dot\q$ and obtain
(\ref{eq:Peq}).

Finally, I turn to my claim that this system has time-reversal invariant
equilibrium correlations $\langle \q(t) \q(0) \rangle$.
One way to see this is from the path integral Lagrangian $L$ of
(\ref{eq:Linertia}), which can be rewritten as
\begin {equation}
   L = {1\over 4T} \, (\ddot\q + \grad V) \g^{-1} 
                      (\ddot\q + \grad V)
       + {1\over4T} \, \dot\q \g \dot\q
       + L_\eta
       + \delta(0) \, \tr \ln e
       + {1\over2T} \, \partial_t \left(\half\dot\q^2 + V\right) .
\end {equation}
As discussed in the main text, the total time derivative at the end
can be thrown away if one is only interested in equilibrium correlations.
The rest of $L$ is manifestly time-reversal invariant, except possibly
the ghost term $L_\eta$.  (It is again important here that my path integrals
are all implicitly defined with symmetric time discretization.)

For this system, the ghost term is actually fairly trivial, as can be seen
by integrating out the ghosts.
Consider the perturbative expansion of the theory, and imagine
an arbitrary ghost loop.  In frequency $\omega$,
the pertrubative ghost propagator is of the form
$[-\omega^2 - i \sigma(0) \omega + M^2]^{-1}$ (where I have suppressed
the indices $i,j$).
The poles in $\omega$ all lie in the lower half complex plane.
Almost any frequency integral corresponding to a ghost loop can then be seen to
vanish by closing the integration contour in the upper half plane.
The one exception is for loops with frequency integrals where the
integrand does not fall off faster than $1/\omega$ as $|\omega| \to \infty$,
since in this case we cannot ignore the contribution of the contour at
infinity.  This situation arises only for ghost loops with a single
ghost propagator and a single vertex of the form
$-i \omega \, \delta\sigma(\q)$.  The resulting frequency integral yields an
effective interaction among the $\q$'s of
\begin {equation}
   L_{\eta} \to - \int_{-\infty}^{+\infty} {d\omega\over 2\pi} \>
      {-i \omega \, \delta\sigma \over
         [-\omega^2 - i \omega \, \sigma(0) + M^2]}
   = - \half \, \delta\sigma .
\end {equation}
Restoring indices $i,j$ and not worrying about keeping track of additive
constants in the action, this becomes,
\begin {equation}
   L_{\eta} \to - \half \tr \sigma .
\end {equation}
This is clearly time-reversal invariant.


\section{Integrating by parts with symmetric discretization}
\label{app:parts}

Consider a term in the action of the form
\begin {equation}
   I = \int dt \> \dot \q \cdot \grad f(\q) ,
\end {equation}
which can be naively integrated to yield a boundary term.  I will review
how this naive integration is justified for the symmetric time discretization
used in this paper.  In that discretization, $I$ really represents
\begin {equation}
   I = \sum_\tau (\q_{\tau+1}-\q_\tau) \cdot
        \grad f\!\left(\q_{\tau+1}+\q_\tau \over 2\right) ,
\end {equation}
where $\tau = t/\Delta t$ is an integer parameterizing the time steps.
Now consider the Taylor expansion of $f(\q_{\tau+1})-f(\q_\tau)$ about
$\bar\q \equiv (\q_{\tau+1}+\q_\tau)/2$:
\begin {equation}
   f(\q_{\tau+1})-f(\q_\tau)
   = \Delta\q \cdot \grad f(\bar\q) + O[(\Delta q)^3] ,
\end {equation}
where $\Delta\q \equiv \q_{\tau+1}-\q_\tau$.  So we can rewrite
\begin {equation}
   I = \sum_\tau \left\{ f(\q_{\tau+1}) - f(\q_\tau) + O[(\Delta q)^3] \right\}
   .
\label{eq:I}
\end {equation}
In a path integral, contributions to the action survive in the
$\Delta t \to 0$ limit if they contribute
$O(\Delta t)$ or more per time step.%
\footnote{
  As a quick mnemonic, think about the contributions of a potential
  term: $\int dt \> U(\q) = \Delta t \sum_t U(\q)$.  Such contributions
  are manifestly $O(\Delta t)$ per time step.
}
The kinetic term determines the size of $\Delta q$ to be $O[(\Delta
t)^{1/2}]$.
Therefore, the $O[(\Delta q)^3]$ term in (\ref{eq:I}) can safely be ignored,
and what's left trivially cancels between successive time steps, except
for boundary terms.

The fact that the error was $O[(\Delta q)^3]$ and not $O[(\Delta q)^2]$
(which would not be ignorable) in (\ref{eq:I}) depended crucially
on the symmetric discretization $\bar\q = (\q_{\tau+1}+\q_{\tau})/2$.


\section{Euclidean Schr\"odinger equation for \boldmath$L_2$}
\label{app:H}

We can directly determine the equilibrium distribution generated by the
generic Lagrangian (\ref{eq:Lgeneric}) by transforming the path integral
into a Euclidean Schr\"odinger equation.
First, recast the path integral over $\q(t)$ as a path integral over
$\q(t)$ and momentum $\p(t)$:
\begin {equation}
   Z_2 = \lim_{\Delta t \to 0} N' \int 
      \left[ \prod_\tau d\p_\tau d\q_\tau\right]
      e^{-S_2(\p,\q)} ,
\end {equation}
\begin {equation}
   S_2(\p,\q) = \sum_\tau \left\{
          -i \p_\tau \cdot (\q_\tau - \q_{\tau-1})
           + \Delta t \, H_2\left(\p_\tau, {\q_\tau + \q_{\tau-1}\over 2}
                          \right) \right\} ,
\end {equation}
\begin {equation}
   H_2(\p,\q) = T \p \, \g^{-1}(\q) \, \p + u(\q) ,
\end {equation}
\begin {equation}
   u \equiv U - {1\over\Delta t} \tr\ln e ,
\end {equation}
which can be checked simply by doing the Gaussian integration over $\p$.
In this form, the path integral is well known to correspond to a
Schr\"odinger equation
\begin {equation}
   [ H_2(\p,\q) ]\W \psi(\q,t) = - \dot \psi(\q,t) ,
\label {eq:Sch1}
\end {equation}
where the subscript W indicates Weyl ordering of the operators $\p$
and $\q$.  The Weyl ordering formula we need in the case at hand
is that%
\footnote{
   See, for example, ref.\ \cite{path} for a review of this fact in the
   present context.
}
\begin {equation}
   [p_i p_j \, A(\q)]\W
   = {1\over4} \left\{ \hat p_i , \left\{ \hat p_j , A(\hat\q) \right\}
                  \right\} ,
\end {equation}
where I've now introduced hats to emphasize that $\p$ and $\q$ are operators.
Taking $\hat\p = -i \grad$,
the Schr\"odinger equation (\ref{eq:Sch1}) then becomes
\begin {equation}
   \dot\psi = \left\{
         T \left[ (\sigma^{-1})_{ij} \nabla_i\nabla_j
              + (\sigma^{-1})_{ij,i} \nabla_j
              + {1\over4} (\sigma^{-1})_{ij,ij} 
         \right]
         - u \right\} \psi .
\label{eq:Sch2}
\end {equation}

I have used the symbol
$\psi$ instead of $P$ in this equation because the equilibrium
result for $\psi$ must be the square root of the probability distribution.
That is, we want the Schr\"odinger equation (\ref{eq:Sch2}) to have
\begin {equation}
    \psi_{\rm eq} = e^{-V(\q)/2T}
\label{eq:psieq}
\end {equation}
as its solution rather than $\exp(-V/T)$.
To understand this, consider the original Lagrangian $L$ from the discussion
of the path integral form of Langevin equations.
$L$ differed from $L_2$
only by a total time derivative.  The Schr\"odinger equation
corresponding to $L$
is simply the Fokker-Planck equation (\ref{eq:FP}), which I'll now write
in the form
\begin {equation}
   \dot P = - \hat H_1 P .
\end {equation}
Both formulations, in terms of $L$ or $L_2$, should generate the same
equilibrium physics---that is, the same time-dependent correlation
functions.  A crucial difference between $\hat H_2$ and $\hat H_1$, however, is
that the operator $\hat H_2$ is Hermitian, while $\hat H_1$ is not.
Now think about what it means to write down a path integral expression for
the equilibrium probability distribution
$\Peq(\tilde\q) = \langle \delta(\q-\tilde\q) \rangle$
in terms of actions that run between
arbitrarily large times $-\T$ and $+\T$.  The corresponding
object in the Schr\"odinger formulation is
\begin {equation}
   \Peq(\tilde\q) = \lim_{\T \to \infty}
   { \langle \q(+\T) | e^{- \hat H \T} | \tilde\q \rangle
     \langle \tilde\q | e^{- \hat H \T} | \q(-\T) \rangle
     \over
    \langle \q(+\T) | e^{- 2 \hat H \T} | \q(-\T) \rangle
   } \,,
\label {eq:Px}
\end {equation}
where $\hat H$ can be either $\hat H_1$ or $\hat H_2$.
What dominates the long-time evolution operator $\exp(- \hat H \T)$
is the equilibrium state, which I'll denote $|\mbox{eq}\rangle$.
The difference between $H_1$ and $H_2$
is that the long-time evolution generated by the
$\hat H_2$ must be
symmetric in its overlap with the initial and final states,
becuase $\hat H_2$ is Hermitian.  That is, in the large $\T$ limit,
\begin {equation}
   \langle \q' | e^{- \hat H_2 \T} | \q'' \rangle \to
   \langle \q' | \mbox{eq}_2\rangle \langle \mbox{eq}_2 | \q'' \rangle .
\end {equation}
$\hat H_1$ is not Hermitian
and so does not have this symmetry.  In fact, we know
from the usual Fokker-Planck equation corresponding to $H_1$ that the
evolution is
dissipative, and the result of long-time evolution is independent of
initial conditions.  So
\begin {equation}
   \langle \q' | e^{- \hat H_1 \T} | \q'' \rangle \to
   \langle \q' | \mbox{eq}_1\rangle .
\end {equation}
In the case of $H_2$,
(\ref{eq:Px}) then becomes
\begin {equation}
   \Peq(\tilde\q) = \Bigl| \langle \tilde\q | \mbox{eq}_2 \rangle \Bigr|^2 ,
\end {equation}
whereas for $H_1$ it becomes
\begin {equation}
   \Peq(\tilde\q) = \langle \tilde\q | \mbox{eq}_1 \rangle .
\end {equation}
One can now see
that the equilibrium wave function represents the square
root of $\Peq$ in the case of $H_2$ but $\Peq$ itself in the case of
$H_1$.

In any case, we can now uniquely
determine $u(\q)$, and hence $U(\q)$, simply by
requiring that the equilibrium amplitude (\ref{eq:psieq}) be a solution of
the Schr\"odinger equation (\ref{eq:Sch2}).
One finds
\begin {equation}
   u = {1\over 4T} (\grad V) \sigma^{-1} (\grad V)
       - {1\over2} \grad(\sigma^{-1}\grad V)
       + {T\over 4} (\sigma^{-1})_{ij,ij} ,
\end {equation}
which precisely reproduces the result (\ref{eq:L2}) for the $L_2$
that describes the Langevin equation, with my claimed result
(\ref{eq:Fstrat}) for $\bFstrat$.


\end {document}